\documentclass[a4paper,fleqn,usenatbib]{mnras}

\usepackage{newtxtext,newtxmath}

\UseRawInputEncoding

\usepackage[T1]{fontenc}
\usepackage{ae,aecompl}
\usepackage{pdflscape}
\usepackage[figuresright]{rotating}
\usepackage{tablefootnote}
\usepackage{longtable}
\usepackage{threeparttable}
\usepackage{caption}
\usepackage{comment}
\usepackage{adjustbox}

\usepackage{graphicx}	
\usepackage{amsmath}	
\usepackage{amssymb}	
\usepackage[LGRgreek]{mathastext}

\title[Spiral and streamer towards a candidate proto-BD]{Observations of spiral and streamer on a candidate proto-brown dwarf}

\author[Riaz, Stamatellos, Machida]{
Riaz, B.,$^{1}$\thanks{E-mail: briaz@usm.lmu.de}
Stamatellos, D.,$^{2}$
Machida, M. N.$^{3}$
\\
$^{1}$  Universit\"{a}ts-Sternwarte M\"{u}nchen, Ludwig Maximilians Universit\"{a}t, Scheinerstra$\beta$e 1, 81679 M\"{u}nchen, Germany
\\
$^{2}$  Jeremiah Horrocks Institute for Mathematics, Physics \& Astronomy, University of Central Lancashire, Preston PR1 2HE, UK
\\
$^{3}$  Department of Earth and Planetary Sciences, Faculty of Sciences, Kyushu University, Fukuoka, Japan
}

\date{Accepted XXX. Received YYY; in original form ZZZ}

\pubyear{2024}

\begin{document}
\label{firstpage}
\pagerange{\pageref{firstpage}--\pageref{lastpage}}
\maketitle

\begin{abstract}


Spirals and streamers are the hallmarks of mass accretion during the early stages of star formation. We present the first observations of a large-scale spiral and a streamer towards a very young brown dwarf candidate in its early formation stages. These observations show, for the first time, the influence of external environment that results in asymmetric mass accretion via feeding filaments onto a candidate proto-brown dwarf in the making. The impact of the streamer has produced emission in warm carbon-chain species close to the candidate proto-brown dwarf. Two contrasting scenarios, a pseudo-disk twisted by core rotation and the collision of dense cores, can both explain these structures. The former argues for the presence of a strong magnetic field in brown dwarf formation while the latter suggests that a minimal magnetic field allows large-scale spirals and clumps to form far from the candidate proto-brown dwarf.

\end{abstract}


\section{Introduction} 
\label{intro}

Early star formation is a chaotic and dynamic process wherein complex gas structures and kinematics are observed. During this stage, thermal and turbulent pressure, magnetic fields, and rotation play an important role during the gravitational collapse of a molecular cloud core to a pre-stellar core and later a protostar. The environment plays an equally important role. Pre-stellar cores are not spherically symmetric but rather tend to form within filamentary structures and continue to accrete from their surroundings via these feeding filaments. Recent surveys with ALMA have led to serendipitous discoveries of asymmetric, spatially extended spirals and streamers that trace such infalling channels, which in some cases are connected to the central disks (e.g., Lee et al. 2023; Yen et al. 2019).

Whether the formation of brown dwarfs is a scaled-down version of low-mass star formation or not still remains an enigma. A test of this hypothesis requires high-sensitivity and high-angular observations of brown dwarfs during their earliest formation stages. Ser-emb 16 provides an ideal case for such a study. This is a young stellar object (YSO) located in the northern tip of the Serpens filament (Gong et al. 2021). It has been classified as a Class 0/Stage 0 YSO (Riaz \& Thi 2022a), with a bolometric luminosity of 0.06$\pm$0.02 L$_{\odot}$ and total (dust+gas) mass of 17.7$\pm$3.2 M$_{Jup}$ (Sect.~\ref{obs};~\ref{alma-obs}), which are well below the sub-stellar limits of $<$0.08 M$_{\odot}$ and $<$0.1 L$_{\odot}$ typically considered in brown dwarf evolutionary models (Burrows et al. 2001). This YSO can thus be considered as a candidate proto-brown dwarf. Previous molecular line surveys have identified Ser-emb 16 to be chemically rich, with strong emission in several high- and low-density molecular species, most notably, mono-deuterated methane and the fundamental complex organic molecule methanol (Riaz \& Thi 2022abc; Riaz et al. 2018; 2019a; 2023). With molecular abundances as high as measured in low-mass protostars, Ser-emb 16 has all the ingredients to probe the initial stages of brown dwarf formation.

\section{Target and ALMA Observations}
\label{obs}

Ser-emb 16 [SSTc2d J182844.8+005126; RA (J2000) = 18h28m44.78s, Dec (J2000) = 00d51m25.79s] is classified as a Class 0/Stage 0 YSO, as described in detail in (Riaz et al. 2018; Riaz \& Thi 2022a). It is located in the Serpens star-forming region [$d$=440$\pm$9 pc; (Ortiz-Leon et al. 2017)]. The bolometric luminosity, L$_{bol}$, is 0.06$\pm$0.02 L$_{\odot}$. The internal luminosity, L$_{int}$, is estimated to be $\sim$0.045$\pm$0.015 L$_{\odot}$. Both L$_{bol}$ and L$_{int}$ are below the luminosity threshold typically considered between very low-mass stars and brown dwarfs ($<$ 0.1 L$_{\odot}$) in brown dwarf evolutionary models (e.g., Burrows et al. 2001). A detailed discussion on the present and final mass of Ser-emb 16 is presented in Sect.~\ref{mass-epoch}.

We note that Pokhrel et al. (2023) have included the Herschel PACS and SPIRE data points in the SED for Ser-emb 16 and have derived a $\sim$5 times higher L$_{bol}$ than our measured value of 0.06 L$_{\odot}$. The Herschel PACS and SPIRE images are at a large angular resolution of 6$\arcsec$-36$\arcsec$ ($>$2600 au), which includes the spiral/streamer seen in our ALMA observations (Sect.~\ref{alma-obs}). The actual size of the source Ser-emb 16 as measured here in ALMA continuum at a $\sim$0.4$\arcsec$ angular resolution is about 1$\arcsec$ or $\sim$588 au (Table~1; Sect.~\ref{alma-obs}). Therefore, the Herschel fluxes measured in a $\geq$6$\arcsec$ beam size suffer from both beam dilution of the compact source as well as contamination from the surrounding cloud material. Due to this, the L$_{bol}$ derived for Ser-emb 16 from the SED that includes the Herschel data points would be incorrect. Note that the classification of this target as Class 0/Stage 0 are based on both the 2-24 $\mu$m SED slope and the HCO$^{+}$ line flux, as discussed in Riaz et al. (2018; 2022a), and are not affected by the Herschel fluxes.

We observed Ser-emb 16 with ALMA in the HCO$^{+}$ (3-2) line at 267.567 GHz in Band 6. The observations were obtained in Cycle 8 (PID: 2021.1.00134.S) in September, 2022. The primary beam of the ALMA 12m dishes covers a field of $\sim$30 arcsec in diameter. Our chosen configuration was a maximum recoverable scale of $\sim$4 arcsec, and an angular resolution of $\sim$0.4 arcsec. The bandwidth was set to 117.18 MHz, which provides a resolution of 244.14 kHz, or a smoothed resolution of $\sim$0.3-0.4 km s$^{-1}$ at 267.5 GHz. One baseband was used for continuum observations where the spectral resolution was set to 976 kHz and the total bandwidth was 1875 MHz. We employed the calibrated data delivered by the EU-ARC. Data analysis was performed on the uniform-weighted synthesized images using the CASA software. The synthetic beam size of the observations is 0.47$^{\prime\prime}\times$0.38$^{\prime\prime}$ or about 221$\times$179 au at the distance to Serpens. The 1-$\sigma$ rms is $\sim$0.15 mJy beam$^{-1}$ for the continuum image, $\sim$10 mJy beam$^{-1}$ in the HCO$^{+}$ line and $\sim$7 mJy beam$^{-1}$ in the c-HCCCH line.

\section{Data Analysis and interpretation}
\label{alma-obs}

 \begin{figure*}
  \centering      
       \includegraphics[width=2.6in]{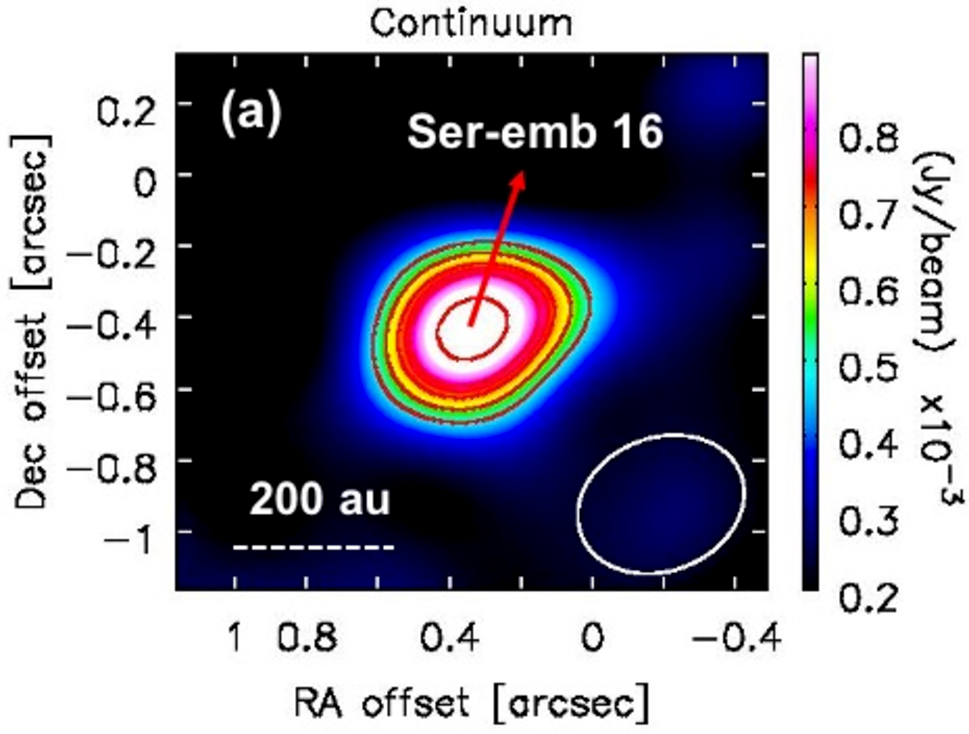}
       \includegraphics[width=2.7in]{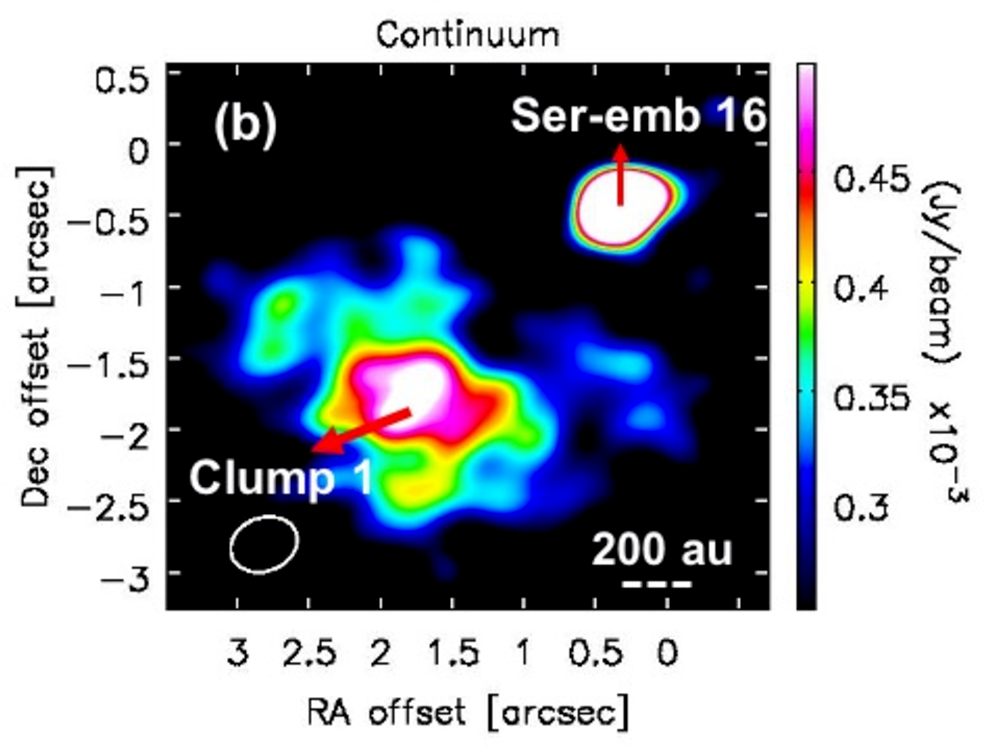}	\\       	
       \includegraphics[width=2.5in]{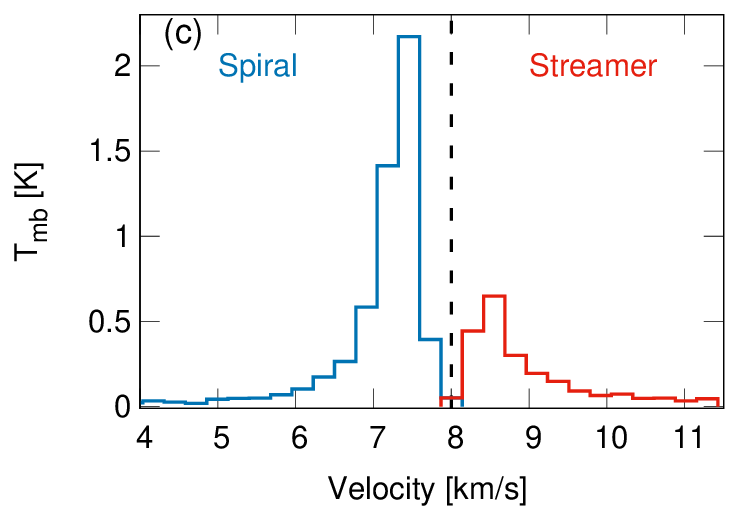}
     \includegraphics[width=2in]{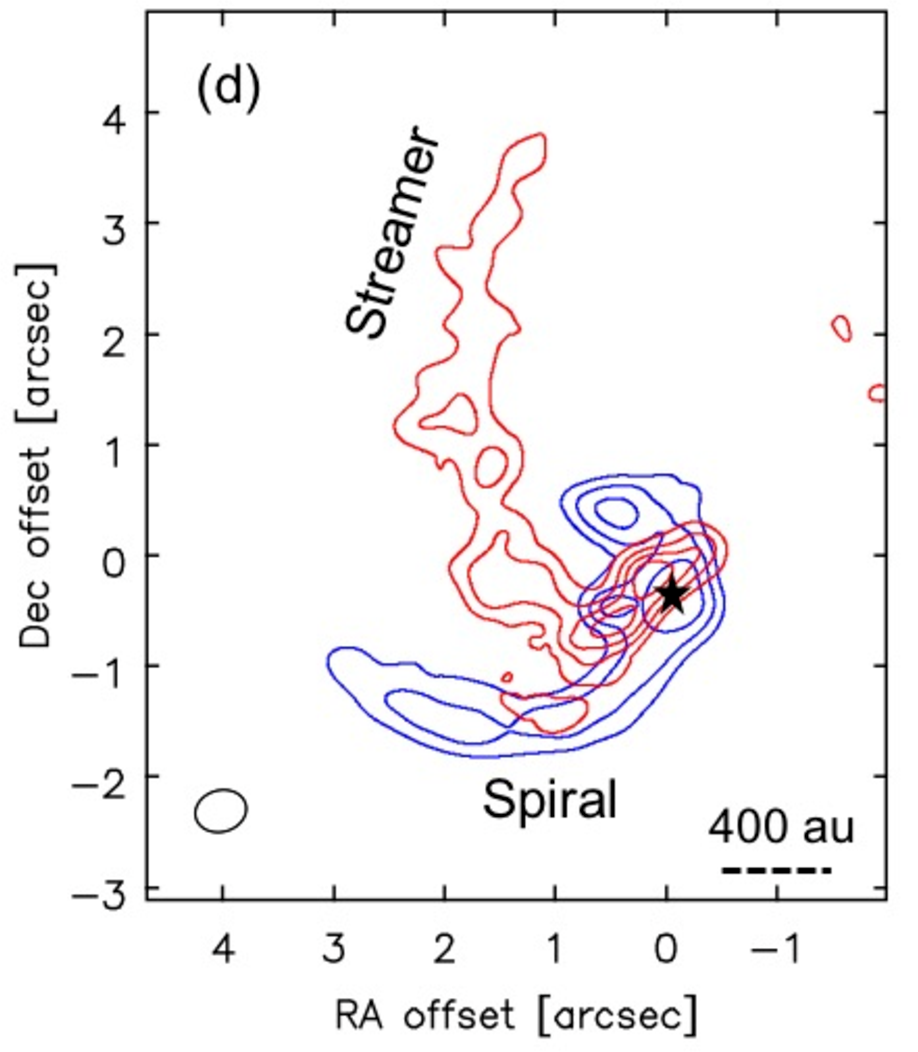}		                 
           \caption{{\bf (a-b)} ALMA 1.16 mm continuum image. {\bf (c)} ALMA HCO$^{+}$ (3-2) spectrum. Dashed line marks the V$_{lsr}$ for Ser-emb 16. {\bf (d)} The moment 8 maps of the blue-shifted spiral and red-shifted streamer overplotted in positions. The contours are from 2-$\sigma$ to 10-$\sigma$ in steps of 2-$\sigma$. The 1-$\sigma$ rms is $\sim$10 mJy beam$^{-1}$. Ser-emb 16 continuum position is marked by a star. }
           \label{obs1}
  \end{figure*}

 \begin{figure*}
  \centering      
       \includegraphics[width=3in]{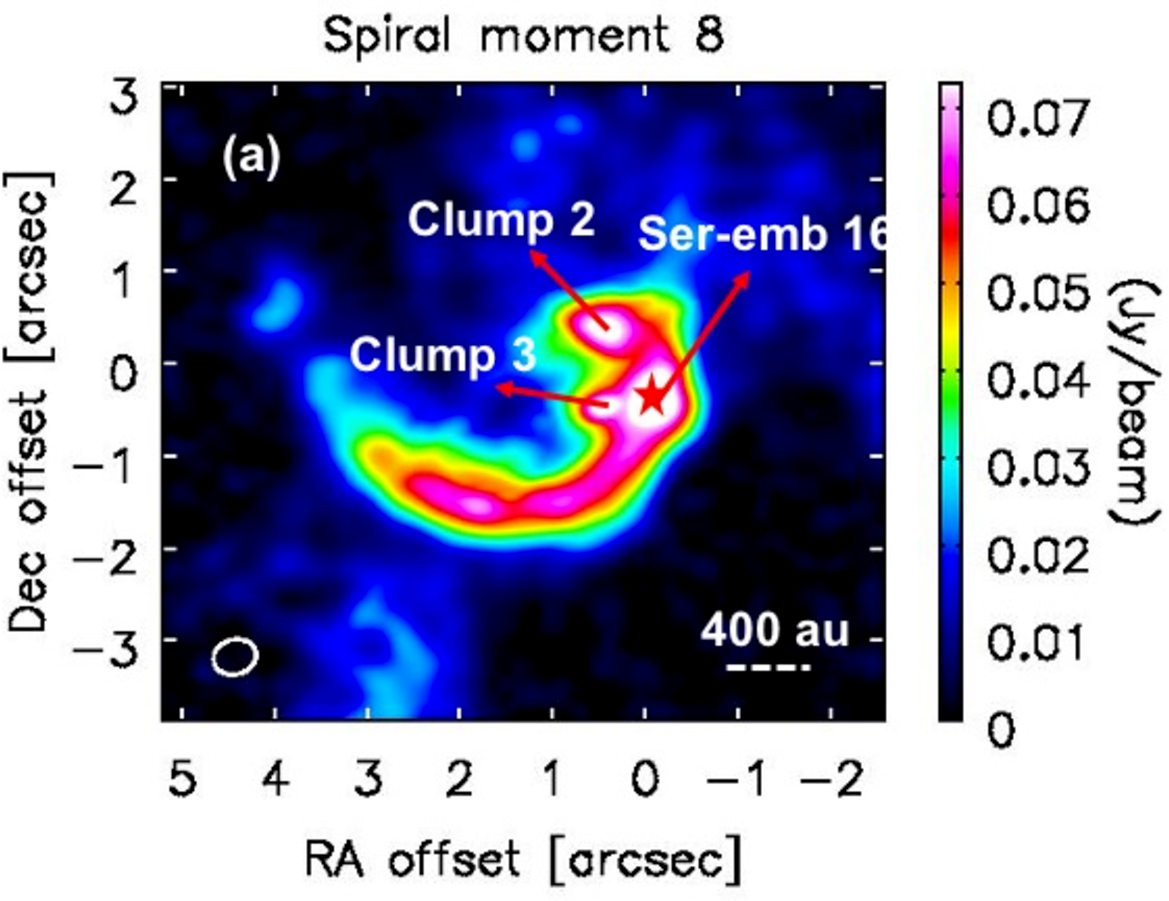} 
       \includegraphics[width=3.1in]{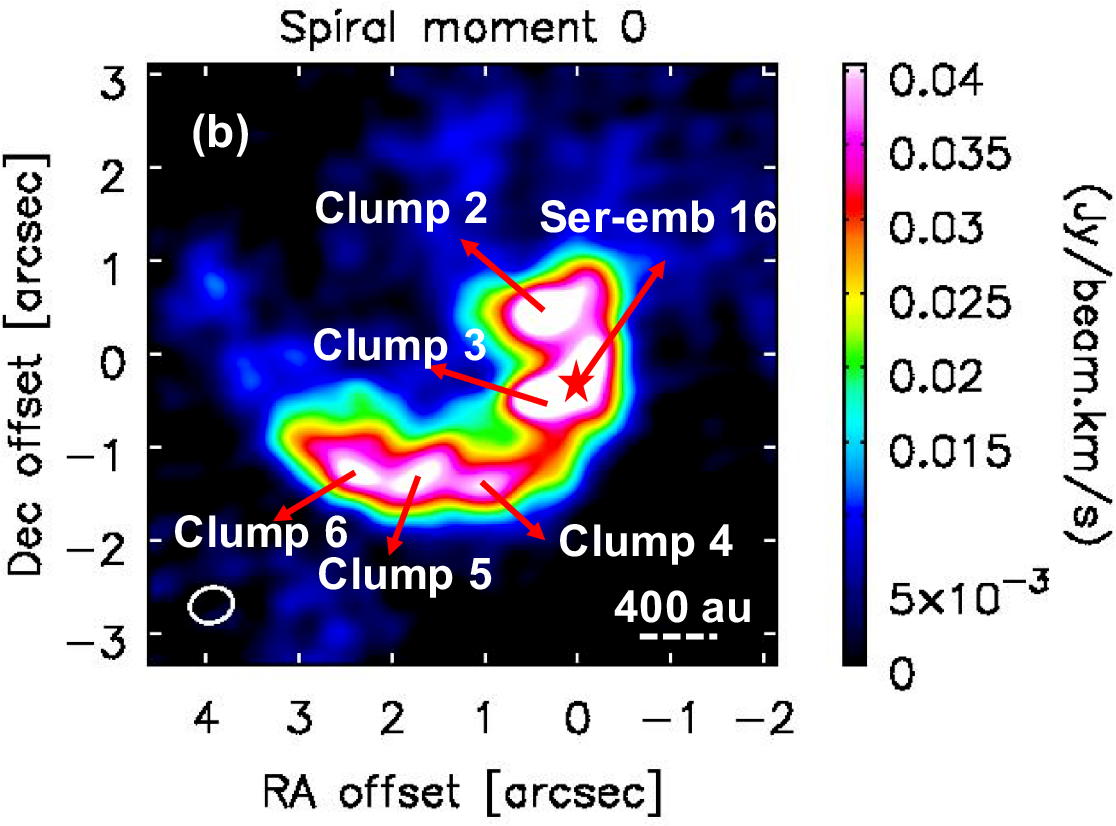}	
       \includegraphics[width=2.9in]{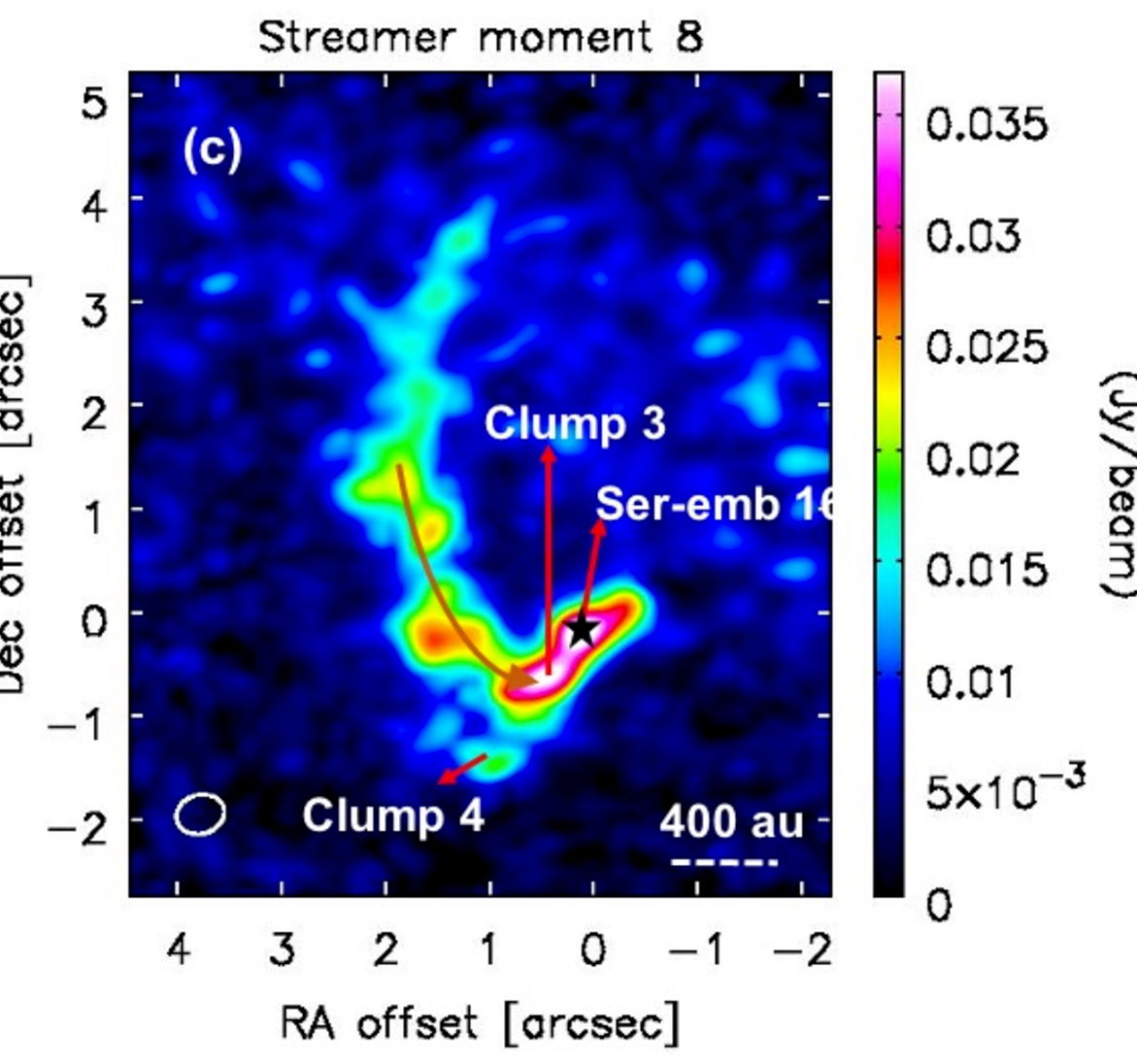}                         
       \includegraphics[width=3.12in]{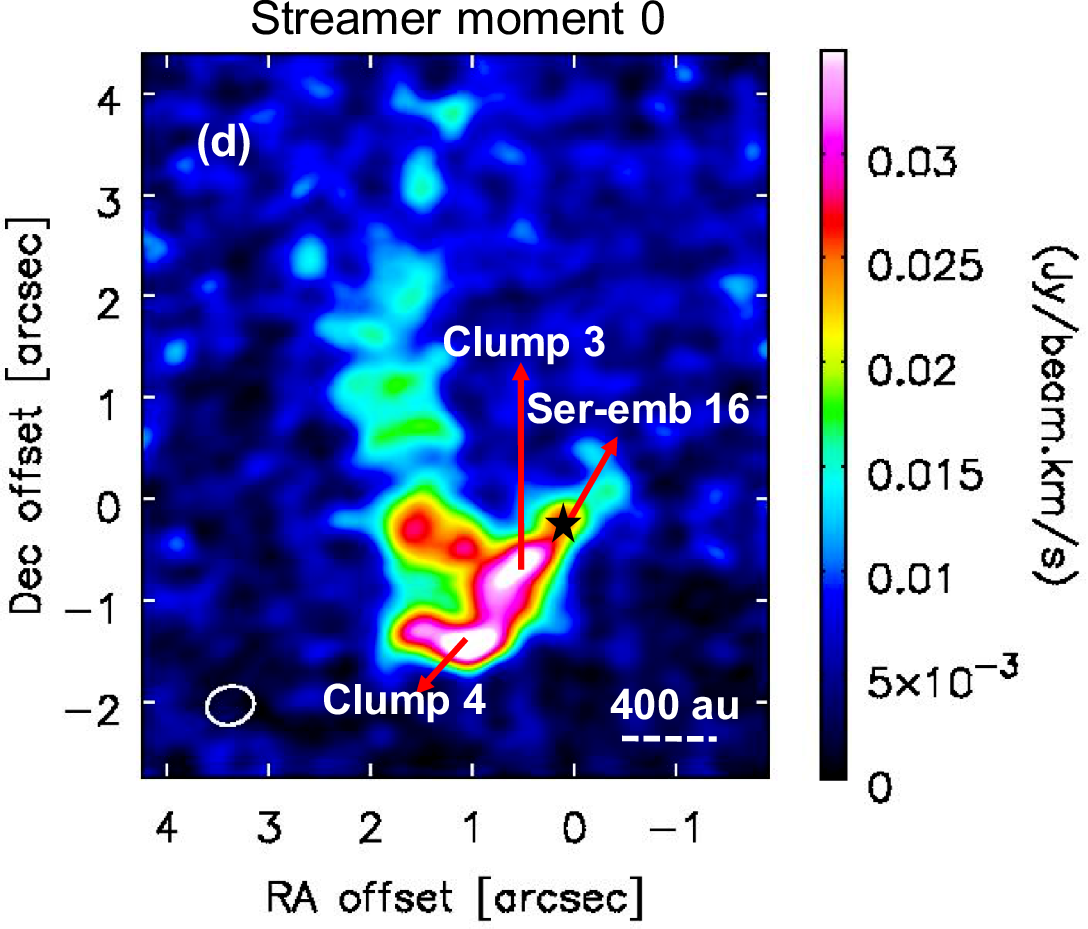}    
       \includegraphics[width=2.9in]{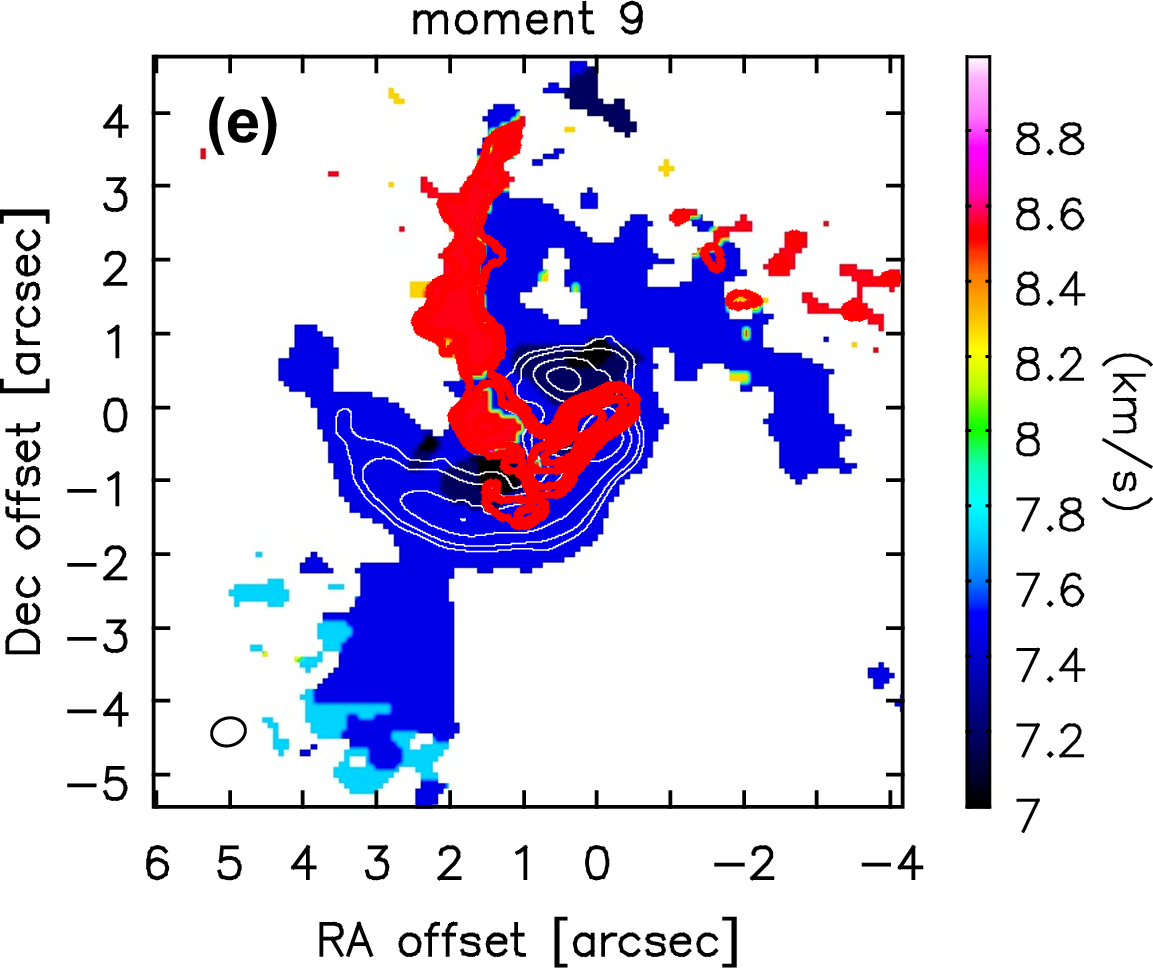}	
           \caption{Moment maps in the HCO$^{+}$ (3-2) line observations. {\bf (a)} Moment 8 map for the spiral. {\bf (b)} Moment 0 map for the spiral. {\bf (c)} Moment 8 map for the streamer. {\bf (d)} Moment 0 map for the streamer. Ser-emb 16 continuum position is marked by a star. {\bf (e)} Moment 9 map for the spiral and streamer. White contours are the spiral structure seen in the moment 0 and 8 maps.}
           \label{obs2}
  \end{figure*}

Figure~\ref{obs1}a shows a bright source detection for Ser-emb 16 in the ALMA 1.16 mm dust continuum map. Ser-emb 16 is resolved in the continuum and has a projected size of 588$\pm$97 au. In addition, we see a fainter peak (labelled `clump 1') at a $\sim$2$^{\prime\prime}$ ($\sim$880) separation from Ser-emb 16, with scattered dust emission around it (Fig.~\ref{obs1}b). Clump 1 is not a resolved point source in the continuum and therefore unlikely (at this stage) to be a companion of Ser-emb 16. It is also difficult to observationally confirm common proper motion pairs that are widely separated at $\sim$2000 au (e.g., Softich et al. 2022).

Figure~\ref{obs1}b shows the observed HCO$^{+}$ (3-2) spectrum. We see two distinct peaks at $\sim$7.4 and $\sim$8.5 km s$^{-1}$ that are mirrored at roughly about the systemic velocity of Ser-emb 16, V$_{lsr}$ = 8.0$\pm$0.2 km s$^{-1}$ (Riaz et al. 2019a). We create moment 0 (integrated value of the spectrum) and moment 8 (maximum value of the spectrum) maps separately over the blue-shifted ($<$8.0 km s$^{-1}$) and red-shifted ($>$8.0 km s$^{-1}$) velocities. A bright arc-like structure termed  ``spiral'' and a fainter elongated structure termed ``streamer'' are observed at different orientations around Ser-emb-16, as seen in Fig.~\ref{obs1}c. The spiral is seen at velocities of $\sim$7.2-7.5 km s$^{-1}$, and the streamer at $\sim$8.5-8.7 km s$^{-1}$. With respect to the V$_{lsr}$ for Ser-emb 16, the spiral is blue-shifted while the streamer is red-shifted (Fig.~\ref{obs1}c). The peak intensity in the spiral is $\sim$3 times higher than the streamer (Fig.~\ref{obs1}b).

 \begin{figure}
  \centering      
       \includegraphics[width=2.8in]{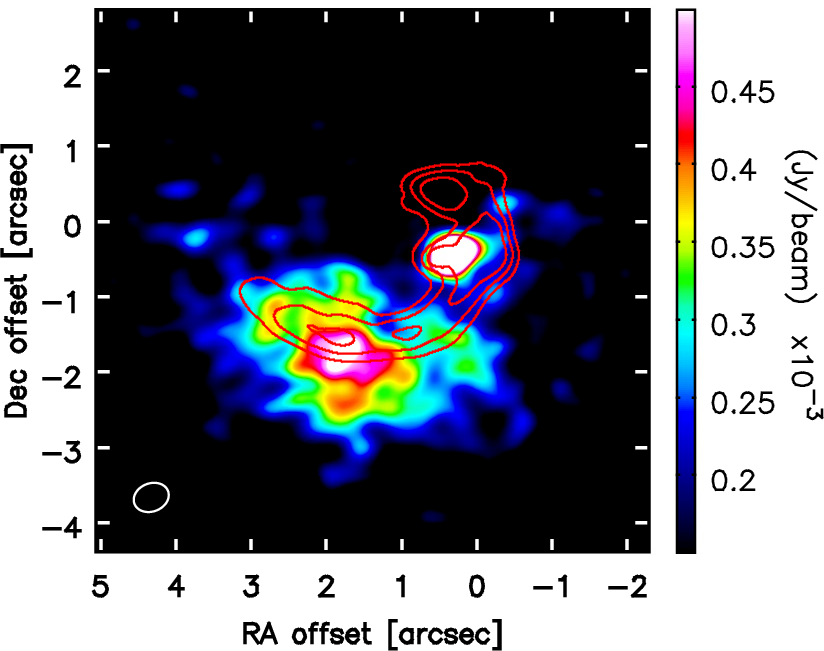}
       \includegraphics[width=2.8in]{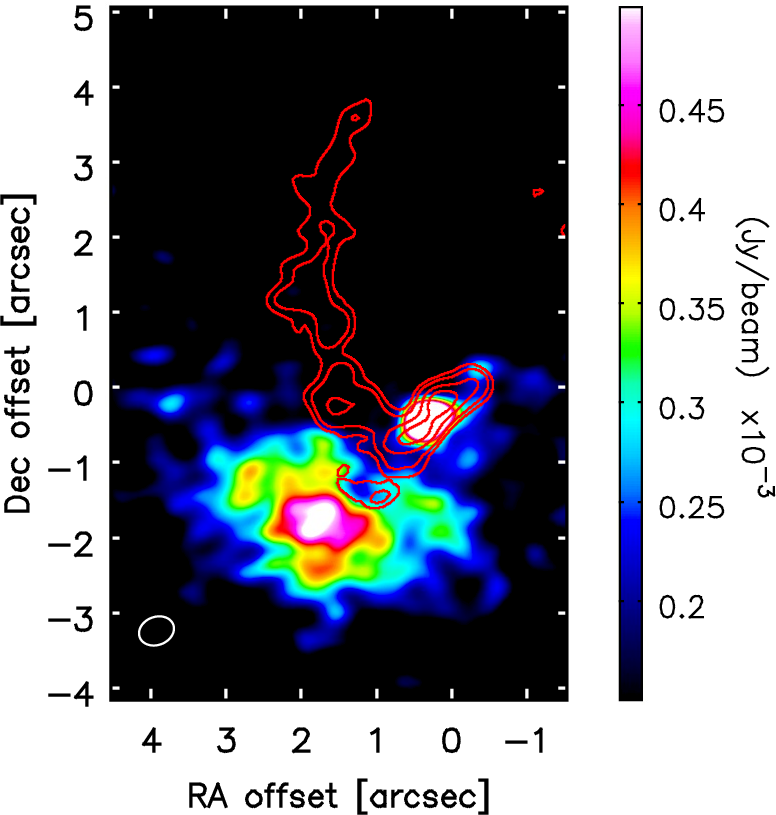}
           \caption{An overlap of the continuum emission (color map) with the spiral (top) and streamer (bottom) moment 8 maps plotted in red contours. The contours are from 2-$\sigma$ to 10-$\sigma$ in steps of 2-$\sigma$. The 1-$\sigma$ rms is $\sim$10 mJy beam$^{-1}$. }
           \label{cont-line}
  \end{figure}

Figure~\ref{cont-line} compares the spiral and streamer moment 8 maps with the dust continuum emission. These structures are not seen in the continuum. The only common feature between the continuum emission and spiral/streamer is the bright peak at the Ser-emb 16 position. If dust and gas are poorly coupled then the morphology observed in dust and gas will be different. The clump 1 seen in the continuum (Fig.~1b) may be related to the small clumps 5,6 seen in the HCO$^{+}$ line image (Fig.~2b). These clumps are likely a part of the large-scale inhomogeneous envelope/pseudo-disk around Ser-emb 16, as suggested by the overlap seen with the spiral in Fig.~\ref{cont-line}. Such inhomogeneous envelopes have been confirmed in recent ALMA studies on proto-brown dwarfs and protostars (e.g., Riaz \& Machida 2021; Shoshi et al. 2023).

Details into the structure of the spiral and streamer can be seen in the moment maps in Fig.~\ref{obs2}. For the spiral, the moment 8 map shows the brightest emission at the position of Ser-emb 16 (Fig.~\ref{obs2}a). In addition, we see a bright clump (labelled `clump 2') $\sim$0.9$^{\prime\prime}$ north of Ser-emb 16 and another small bar-like clump (labelled `clump 3') at $\sim$0.6$^{\prime\prime}$ eastward of it. In comparison, the moment 0 map shows a fragmented spiral with multiple clumps (Fig.~\ref{obs2}b). Clumps 2,3,5,6 are detected at a signal-to-noise ratio (S/N) of $>$5 while Clump 4 has a S/N$\sim$2. The streamer shows an elongated curved structure (Fig.~\ref{obs2}c, d). The peak intensity in the moment 8 map (Fig.~\ref{obs2}c) is seen at the position of Ser-emb 16, with an additional bright peak at a $\sim$0.8$^{\prime\prime}$ separation, close to the clump 3 position seen in the spiral. Several faint blobs are seen along the length of the streamer. The moment 0 map for the streamer (Fig.~\ref{obs2}d) shows a similar curved shape, with a bright clump 4 that was faintly detected in the spiral.

We have used the CASA tasks {\it uvmodelfit} and {\it imfit} to measure the source size, continuum and line fluxes by fitting one or more elliptical Gaussian components on the continuum and HCO$^{+}$ (3-2) line images. Using the HCO$^{+}$ integrated line flux and the source size, the HCO$^{+}$ column density, H$_{2}$ gas mass and H$_{2}$ number density were derived following the method described in Riaz et al. (2018). These parameters are listed in Table~\ref{pars} for Ser-emb 16, spiral, and streamer.

\begin{table*}
\centering
\caption{Physical Parameters}
\label{pars}
\begin{tabular}{llllllll} 
\hline

Object 	  & Size~\tnote{a} & HCO$^{+}$ Line Flux& H$_{2}$ gas mass & H$_{2}$ number density \\

\hline

Ser-emb 16 & FWHM~\tnote{b} ~= (1.11$^{\prime\prime} \pm$0.2$^{\prime\prime}$)$\times$(0.78$^{\prime\prime} \pm$0.1$^{\prime\prime}$); PA = 99$^{\circ} \pm$18$^{\circ}$ &  76.43$\pm$0.90 mJy & 17.6$\pm$3.2 M$_{Jup}$ & 2.5$\times$10$^{11}$ cm$^{-3}$ \\
 		& FWHM~\tnote{c} ~= (1.0$^{\prime\prime} \pm$0.2$^{\prime\prime}$)$\times$(0.7$^{\prime\prime} \pm$0.2$^{\prime\prime}$); PA = 97$^{\circ} \pm$18$^{\circ}$	 & & & \\  
		   & Projected ~= 588$\pm$97 au  & & & \\ 
\hline
Spiral	   & FWHM~\tnote{b}~~\tnote{c} ~= (3.80$^{\prime\prime} \pm$0.16$^{\prime\prime}$)$\times$(2.80$^{\prime\prime} \pm$0.10$^{\prime\prime}$); PA = 119.8$^{\circ} \pm$2.0$^{\circ}$ &  48.23$\pm$1.10 mJy & 0.094$\pm$0.01 M$_{\odot}$ & 3.6$\times$10$^{10}$ cm$^{-3}$ \\
		   & Projected ~= 2077$\pm$83 au  	 & & & \\
\hline
Streamer	   & FWHM~\tnote{b}~~\tnote{c} ~= (5.82$^{\prime\prime} \pm$0.48$^{\prime\prime}$)$\times$(3.10$^{\prime\prime} \pm$0.18$^{\prime\prime}$); PA = 16.2$^{\circ} \pm$2.3$^{\circ}$ &  32.02$\pm$1.20 mJy & 0.045$\pm$0.01 M$_{\odot}$ & 7.2$\times$10$^{9}$ cm$^{-3}$ \\
		   & Projected ~= 2901$\pm$225 au  	 & & & \\

\hline
\end{tabular}
\begin{tablenotes}
\item[a] FWHM is the full width at half-maximum. PA is the position angle. The projected size is at the distance to Serpens (440$\pm$9 pc). 
\item[b] The size convolved with the beam. 
\item[c] The size deconvolved with the beam. 
\end{tablenotes}
\end{table*}

Ser-emb 16 is detected in the continuum, with an integrated continuum flux of 3.22$\pm$0.63 mJy. Following the method described in Riaz et al. (2019b), the dust continuum mass for Ser-emb 16 is 0.11$\pm$0.02 M$_{Jup}$, assuming a dust temperature of 10 K, and dust opacity at 1.16 mm of 0.02582 cm$^{2}$ g$^{-1}$ for $\beta$ = 1.0 or for optically thin emission. The H$_{2}$ gas mass for Ser-emb 16 is 17.6$\pm$3.2 M$_{Jup}$ (Table~\ref{pars}). This implies a total (dust+gas) mass of 17.7$\pm$3.2 M$_{Jup}$, and a gas to dust mass ratio of 160 for Ser-emb 16. The H$_{2}$ gas and dust number density is 2.5$\times$10$^{11}$ cm$^{-3}$ and 0.44$\times$10$^{5}$ cm$^{-3}$, respectively, for Ser-emb 16. It is much more dense in gas than in the dust.

For the multiple clumps seen in the spiral and streamer, clump 1 is detected in the continuum (Fig.~\ref{obs1}a) but not the HCO$^{+}$ line image, while clumps 2-6 observed in the spiral and streamer (Fig.~\ref{obs2}) are undetected in the continuum image. The integrated continuum flux for clump 1 is 5.97$\pm$0.13 mJy, and the resulting dust mass is 0.20$\pm$0.04 M$_{Jup}$. For the clumps 2-6, the H$_{2}$ gas mass is in the range of 0.01-0.15 M$_{\odot}$, with H$_{2}$ gas number density is in the range of (1.1-6.5)$\times$10$^{11}$ cm$^{-3}$. These are thus high-density clumps that could potentially evolve individually to single brown dwarfs or may form a multiple system of sub-stellar/very low-mass objects.

Figure~\ref{obs2}e shows the velocity structure for the spiral and the streamer in a moment 9 (coordinate of the maximum value of the spectrum) map. While the velocity field is smooth, we do not see any clear velocity gradient for either of these structure over their narrow velocity range. The velocity is nearly constant at $\sim$7.4 km s$^{-1}$ for the spiral and $\sim$8.5 km s$^{-1}$ for the streamer. This makes it difficult to constrain the kinematics of these structures. The velocity resolution of ALMA observations is $\sim$0.3-0.4 km s$^{-1}$. The enclosed mass at a radius is used to estimate the free-fall velocity. For the mass and size of Ser-emb 16, spiral, and streamer (Table~\ref{pars}), the free-fall velocity is $\sim$0.2-0.4 km s$^{-1}$. Velocity shifts of $<$0.6-0.8 km s$^{-1}$ cannot be resolved in the observations. On the other hand, the fact that there is no clear velocity dispersion is proof of the existence of a very low-mass object or a proto-brown dwarf since the velocity dispersion, which is (assumed to be) caused by the free-fall velocity, would be higher for a more massive object.

Typically, blue-shifted emission in HCO$^{+}$ is considered a sign of infalling material while red-shifted emission is associated with outflowing gas (e.g., Riaz \& Machida 2021). The PA of 16$^{\circ}$ for the streamer is nearly perpendicular to the PA of 100$^{\circ}$ for Ser-emb 16. Assuming that the PA for Ser-emb 16 is tracing the (pseudo-)disk, this suggests that the streamer could be an outflow. However, there is no detection of a molecular outflow in the CO (2-1) line (Sect.~\ref{outflow}). Simulations have shown that if gas accretes onto cores along accretion channels or feeding filaments observed as spirals/streamers then accretion is not uniform and some parts of the gas may flow in the ``outward'' direction (e.g., Kuffmeier et al. 2017; Bate et al. 2014). Depending on the line-of-sight and the orientation of the accretion channel, a streamer can appear to be moving in the ``outward'' direction even when it is infalling material (see for e.g., Fig. 14 of Riaz \& Machida 2021). This could explain the red-shifted streamer versus blue-shifted spiral observed here, both of which are likely tracing accretion channels.

\section{Non-detection of a CO outflow}
\label{outflow}

Interestingly, no molecular outflow is seen in the CO (2-1) line image. There is no CO detection at a $>$2-$\sigma$ level within 10$^{\prime\prime}$ of Ser-emb 16. If an extended molecular outflow is not observed then this could be due to a mis-alignment of the outflow, disk, and local magnetic field (Sect.~\ref{theory}). The directions of the outflow and disk in the misalignment model of Machida et al. (2020) significantly change at different spatial scales, and is also dependent on the velocity of the outflow. There may be an asymmetric small-scale outflow driven by Ser-emb 16 but the direction of the outflow could be aligned in a way that it is hidden from the line of sight. Ser-emb 16 being shrouded in circumstellar material at its early stage makes it difficult to probe a high-velocity jet that can only be detected in certain optical/near-infrared forbidden emission lines (e.g., Riaz \& Bally 2021).

The non-detection in CO line emission could instead be due to a high fraction of CO depletion, which is expected at an early formation stage. Ser-emb 16 has a significant fraction ($\sim$90\%) of CO depleted from the gas phase and frozen on to the surfaces of grains (Riaz et al. 2019a).

\section{Warm Carbon-Chain Species}
\label{c3h2}

 \begin{figure*}
  \centering      
       \includegraphics[width=5in]{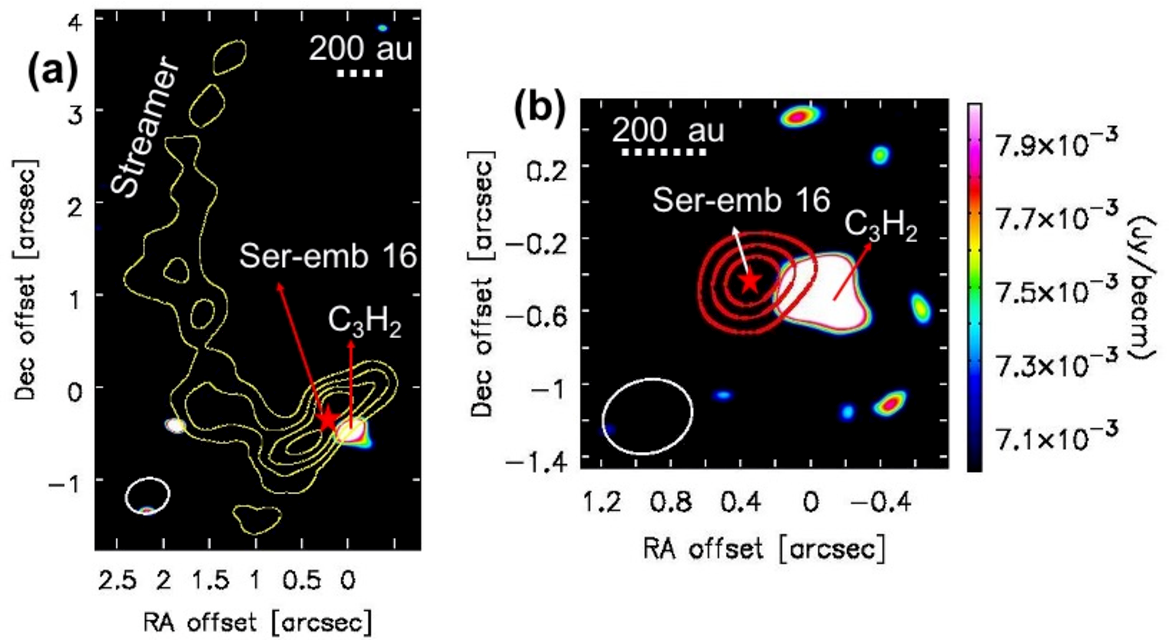} 
           \caption{The moment 8 map in the c-HCCCH line (color raster map) compared with the moment 8 map for the streamer (yellow contours) and the continuum emission (red contours). Red star marks the position for Ser-emb 16. The contours are from 2-$\sigma$ to 10-$\sigma$ in steps of 2-$\sigma$. The 1-$\sigma$ rms is $\sim$10 mJy beam$^{-1}$ in the HCO$^{+}$ line and $\sim$7 mJy beam$^{-1}$ in the c-HCCCH line. }
           \label{c3h2-mom0}
  \end{figure*}

We also report the first detection in the ortho c-HCCCH 5(2, 3)-4(3, 2) transition line for Ser-emb 16. Figure~\ref{c3h2-mom0} shows the moment 8 map for c-HCCCH overlapped with the moment 8 map for the streamer and the continuum map for Ser-emb 16. The curved streamer becomes flattened at its base and appears to be connected to the Ser-emb 16 position. The c-HCCCH emission peaks near the Ser-emb 16 position, slightly shifted from the continuum peak. The impact of the streamer may have produced emission in c-HCCCH close to the candidate proto-brown dwarf (Fig.~\ref{c3h2-mom0}b). This is the first observational evidence of the influence of external accretion that has triggered warm carbon-chain chemistry and affected the chemical composition in the close vicinity of a candidate proto-brown dwarf. There is no c-HCCCH emission seen elsewhere along the spiral or streamer or at the position of any other clumps seen in the HCO$^{+}$ map.

c-HCCCH is considered a warm carbon-chain molecule, the formation of which is triggered by the evaporation of CH$_{4}$ from the grain mantles in a lukewarm ($\sim$20-40 K) and dense (10$^{8}$-10$^{10}$ cm$^{-3}$) region in the close vicinity of a protostar (e.g., Sakai et al. 2008). The ortho c-HCCCH 5(2, 3)-4(3, 2) transition line detected here has an upper state energy level of 41.0 K and a critical density of 1.1$\times$10$^{8}$ cm$^{-3}$. The H$_{2}$ gas number density for Ser-emb 16 is (2.5$\pm$0.1)$\times$10$^{11}$ cm$^{-3}$ (Table~1), much higher than the c-HCCCH critical density. The detection of c-HCCCH suggests that the emission is tracing the warm corino region of the candidate proto-brown dwarf, where the densities are predicted to be $\geq$10$^{8}$ cm$^{-3}$ and the temperature can reach values of $>$20 K (Riaz \& Thi 2022b).

\section{Theoretical interpretations}
\label{theory}

 \begin{figure*}
  \centering    
        \includegraphics[width=2.3in]{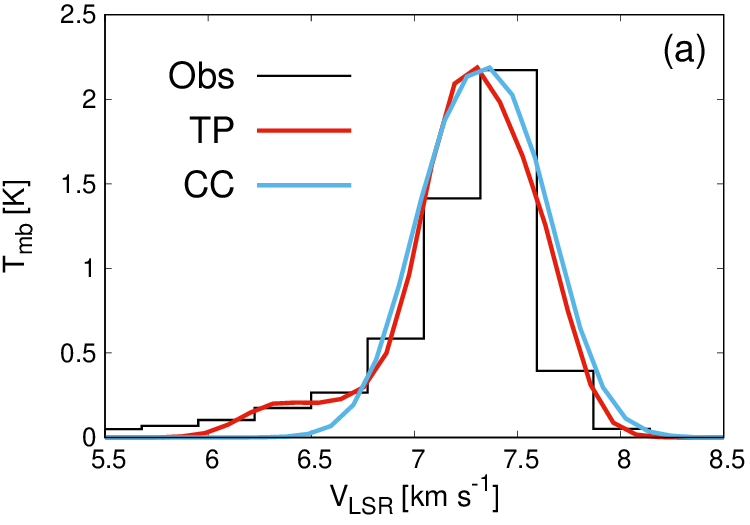}		
        \includegraphics[width=2.36in]{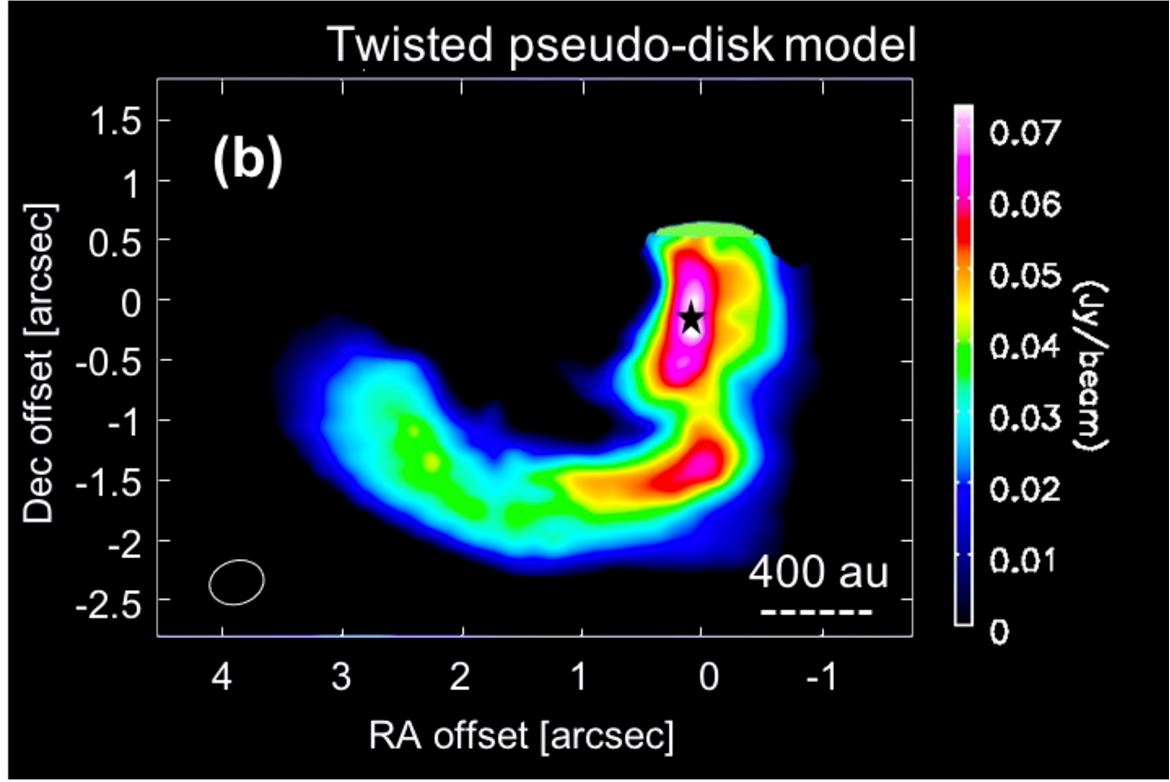}		
        \includegraphics[width=2.21in]{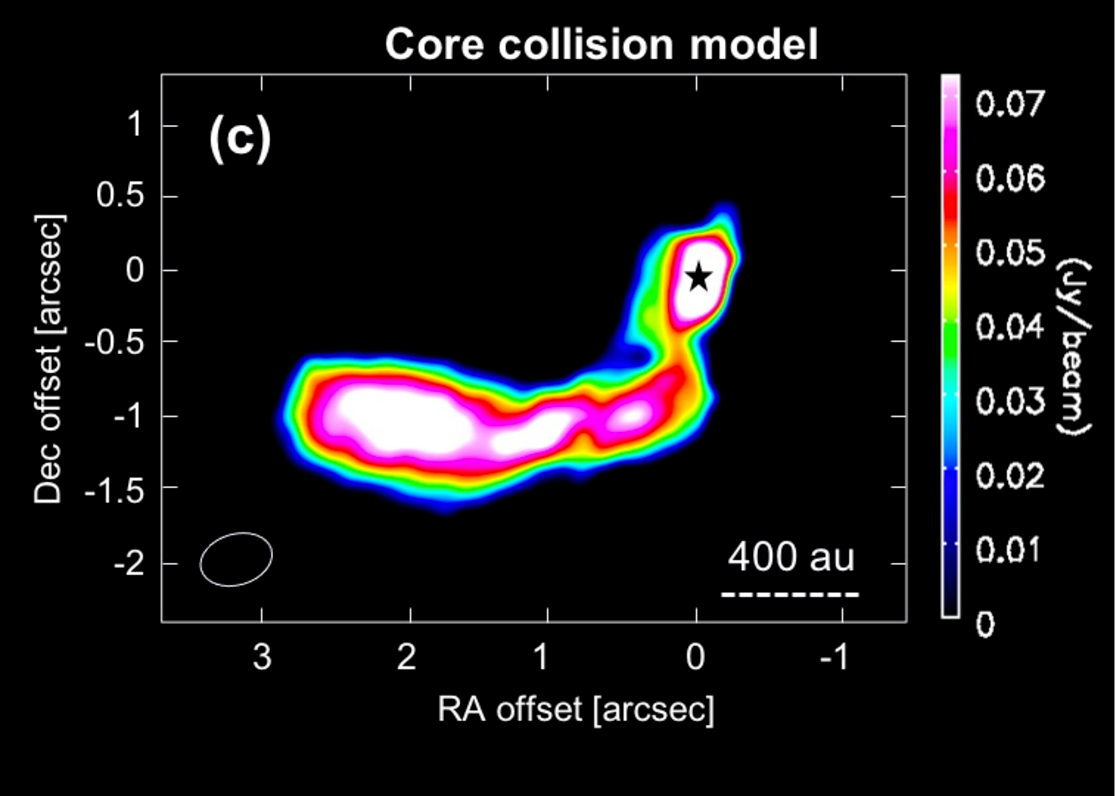}	            
           \caption{{\bf (a)} Observed HCO$^{+}$ (3-2) spectrum for the spiral (black) and the best model fits for the twisted pseudo-disk (TP) model (red) and the core collision (CC) model (blue). {\bf (b-c)} Moment 0 maps from the best match to the observed spiral using the twisted pseudo-disk model and core collision model. Star marks the position of the candidate proto-brown dwarf. }
           \label{model-fit}
  \end{figure*}

While streamers have been observed in low- and high-mass forming stars and explained using various models (e.g., Machida et al. 2009; Bate et al. 2014; Machida et al. 2020), the formation of a large, dense spiral around a candidate proto-brown dwarf is observed for the first time. Here, we explore two different scenarios to explain its possible origin. A detailed explanation of the simulations and the methodology for the line radiative transfer modelling is provided in Appendix A.

We consider the magnetohydrodynamical simulations presented in (Machida et al. 2020) of magnetized rotating clouds in which the rotation axis is misaligned with the global magnetic field by an angle ranging between 0$^{\circ}$ and 90$^{\circ}$. We have improvised these simulations to fit our observations, as explained in Appendix A1. We find a model with an initial misalignment angle of 30$^{\circ}$ best matches the observed structure in the moment 0 map (Fig.~\ref{model-fit}b). The model map shows a similar curved structure with clumps near the candidate proto-brown dwarf position. Modelling indicates that the spiral corresponds to the pseudo-disk structure formed by the process of gravitational collapse with the rotation of an east-west angular momentum vector, which will cause the pseudo-disk to drag the magnetic field and finally connect it to the candidate proto-brown dwarf. In this case, the central magnetic field is twisted due to the rotation of the pre-stellar dense cloud core and directed at 30$^{\circ}$ to the global magnetic field. This model can also explain the lack of an outflow detection in the observations possibly due to the fact that the directions of outflow, disk, and local magnetic field are never aligned during the early accretion stage.

For the second scenario, we consider the formation of a large spiral due to a collision between two small cores, one gravitationally unstable primary core and one stable secondary core, in a young star forming region (Appendix A2). We find that a model with a primary core mass of $\sim$0.2 M$_{\odot}$ and size of $\sim$5000 au, and a secondary core mass of $\sim$0.1 M$_{\odot}$ and a size of $\sim$1000 au best matches the observed structure in the moment 0 map (Fig.~\ref{model-fit}c). The model map shows a similar $\sim$2000 au size of the spiral with multiple bright clumps in it, as seen in the observed map (Fig.~\ref{obs2}b). The spiral and the clumps are a result of the dynamical interaction between the two clouds and the compressed layer that forms at their interface as they collide. The asymmetrical shape of these structures is due to the collision not happening head-on, which is the most likely scenario within star forming regions.

We thus have two contrasting scenarios to explain the observations. If the spiral is the pseudo-disk that initially formed along the magnetic field lines around Ser-emb 16 and has been twisted by the rotation of the core, then this suggests that a strong magnetic field is important in brown dwarf formation. In contrast, in the simulations of core collision, the magnetic field is minimal. This allows large-scale spirals and clumps to form far from the center. Faint streamer-like structures are also produced during these simulations but are dispersed soon after formation (Appendix A).

The asymmetric shapes of the spiral/streamer in the observations and their similar sizes suggest a non-spherical envelope or inhomogenous cloud that could be as large as $\sim$2000-3000 au. If the spiral and streamer are feeding filaments then the free-fall timescale will be $\sim$0.09-0.15 Myr at their $\sim$2000-2800 au spatial scales. This free-fall timescale is comparatively longer than the typical Class 0 lifetime of a normal protostar estimated by statistics to be $\sim$0.05 Myr (e.g., Evans et al. 2009), and suggests a longer timescale of the main accretion phase in brown dwarfs.

Both the spiral and streamer are connected to Ser-emb 16 and form feeding filaments transporting infalling material towards the candidate proto-brown dwarf. In previous works, streamers are considered to be infalling material that has an external origin from an inhomogeneous material outside the star-forming core, while spirals are considered to be infalling material with an internal origin from the envelope or pseudo-disk inside the core (e.g., Lee et al. 2023; Goodwin et al. 2004; Bate et al. 2014; Machida et al. 2020; Yen et al. 2019; Kuffmeier et al. 2017). Based on modelling of the spiral around Ser-emb 16, the twisted pseudo-disk scenario implies that the spiral is infalling material from the circumstellar envelope, while the core collision scenario implies that the spiral is external infalling material from the ambient cloud onto the cores or fragments.

\section{Present and Final Mass}
\label{mass-epoch}

Ser-emb 16 has a total (dust + gas) mass of $\sim$17.7 M$_{Jup}$ derived from the ALMA observations. The central object mass in a Class 0 proto-brown dwarf cannot be directly measured since it is an embedded system and the size of the central object is too small to be resolved even at a $\sim$0.1$\arcsec$ angular resolution. The kinematic estimate of the system (central mass and circumstellar material) using the free-fall or Keplerian velocities is the most plausible way to measure the mass of the central object. However, as discussed in Sect.~\ref{alma-obs}, the velocity resolution of the ALMA observations is not adequate to clearly distinguish infall and Keplerian kinematics. Thus both the angular and velocity resolution are a hindrance to directly measure the central object mass. We can instead estimate the central object mass to be $\leq$10 M$_{Jup}$ at this stage, based on the simulations of brown dwarf formation (Machida et al. 2009). This is typically expected for such very low L$_{bol}$ cases that are in their early Class 0 stage (e.g., Machida et al. 2009; Riaz \& Machida 2021). We thus estimate the total mass of the Ser-emb 16 system (central mass and circumstellar material) of $\leq$28 M$_{Jup}$ in the present epoch. 

It is very difficult to predict the final mass for such a system that is in the main accretion phase. However, observations and theoretical studies (e.g., Machida \& Hosokawa 2013; Machida et al. 2009) have shown that about 70\% of the infalling or envelope mass (including disk and pseudo disk mass) is blown away from the region near the protostar or proto-brown dwarf into interstellar space by the protostellar jet/outflow. 

In such a scenario, if we assume that all of the gas mass in the spiral (0.094$\pm$0.01 M$_{\odot}$) and the streamer (0.045$\pm$0.01 M$_{\odot}$) is accreted onto Ser-emb 16 and about 70\% of this mass is ejected then the final mass is expected to be $\sim$0.05 M$_{\odot}$. Thus, brown dwarf formation is possible for Ser-emb 16 because a significant fraction of the disk and pseudo-disk/envelope mass is expected to be ejected.

\section{Conclusions}

These ALMA observations provide a unique insight into the early formation stages of brown dwarfs and into the role of gas accretion from the external environment. A comparison of the observations with the models support a gravitational infall scenario. This reflects a unique case of a brown dwarf forming in a star-like fashion where the candidate proto-brown dwarf system is in the early formation stage or the main accretion phase and shows spirals and streamers (asymmetric mass accretion), as seen around forming stars.

We have provided two possible theoretical scenarios to interpret the observations and discussed their similarities and differences. Other scenarios may also be possible and further modelling is needed to fit all the details of the current observations, which will be explored in future work.

\section*{Acknowledgements}

B.R. acknowledges funding from the Deutsche Forschungsgemeinschaft (DFG) - Projekt number RI-2919/2-3.

\section{Data Availability}

The ALMA data presented here is from the Cycle 8 program PID: 2021.1.00134.S. This data will become publicly available at the end of the proprietary period. It can then be accessed through the ALMA archives.

\appendix

\section{Models}
\label{models}

\subsection{Twisted spiral model}

We consider the magnetohydrodynamical simulations presented in Machida et al. (2020) of magnetized rotating clouds in which the rotation axis is misaligned with the global magnetic field by an angle ranging between 0$^{\circ}$ and 90$^{\circ}$. 

Machida et al. (2020) studied the contraction of a molecular core which has a uniform magnetic field, rigid-body rotation speed, and density distribution corresponding to a Bonnor-Ebert sphere at an isothermal temperature T = 10 K. In these simulations, the cloud evolution is calculated from the pre-stellar cloud stage until about 500 yr after protostar formation. This timing corresponds to $\sim$6.3$\times$10$^{4}$ yr since the gravitational infall initiated within the dense core. 

The numerical settings used in the present paper are almost the same as in Machida et al. (2020), except changes in the central density n$_{c,0}$ and the density enhancement factor {\it f} of the initial cloud core. In Machida et al. (2020), the initial cloud has a central density n$_{c,0}$ = 6$\times$10$^{5}$ cm$^{-3}$ and a density enhancement factor {\it f} = 1.8, while n$_{c,0}$ = 4$\times$10$^{6}$ cm$^{-3}$ and {\it f} = 1.1 was adopted in this study. Also, the initial cloud has a mass of 2.1 M$_{\sun}$ in Machida et al. (2020), while it has a mass of 0.22 M$_{\sun}$ in this study.

\begin{figure*}
  \centering      
    \includegraphics[width=3.in]{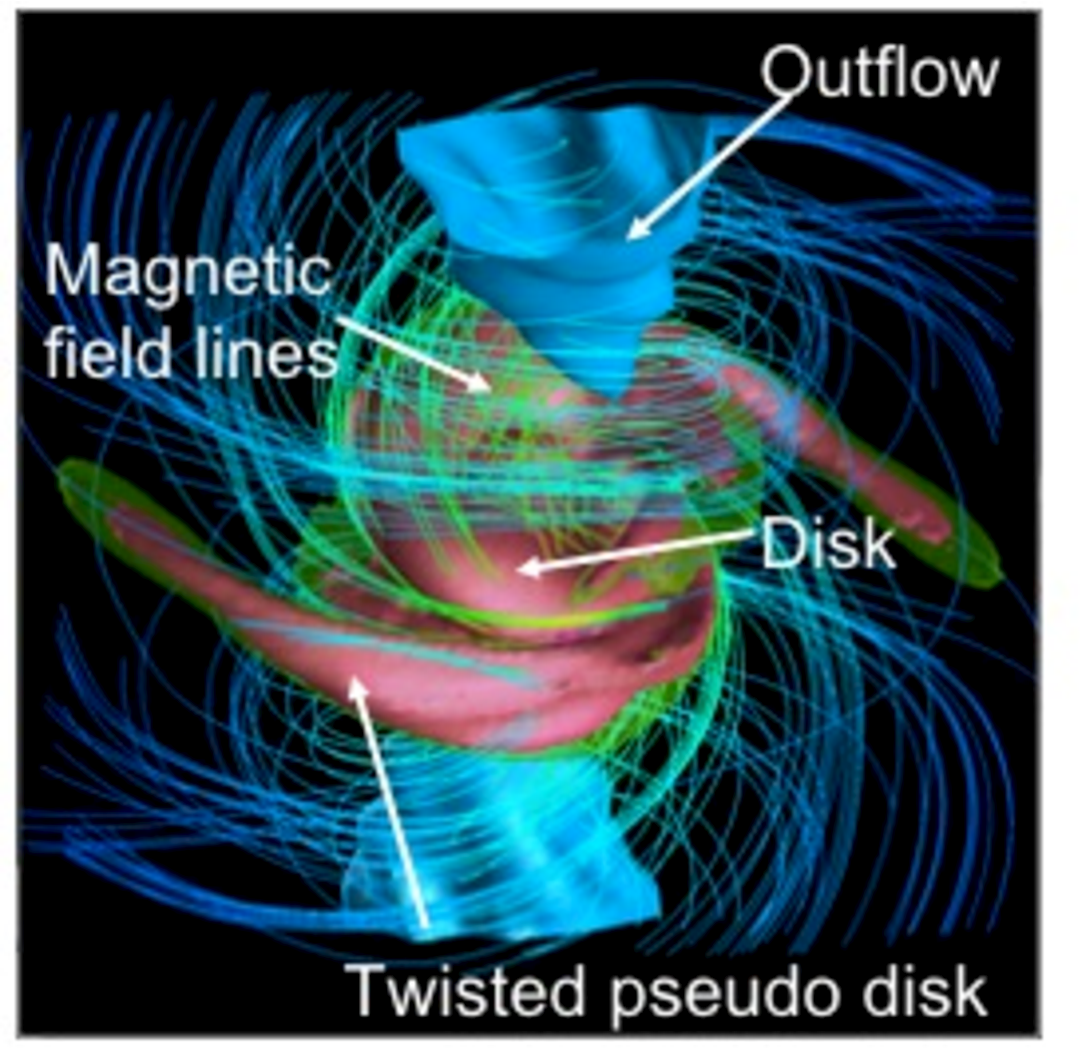} \\	\vspace{0.1in}
    \includegraphics[width=3.3in]{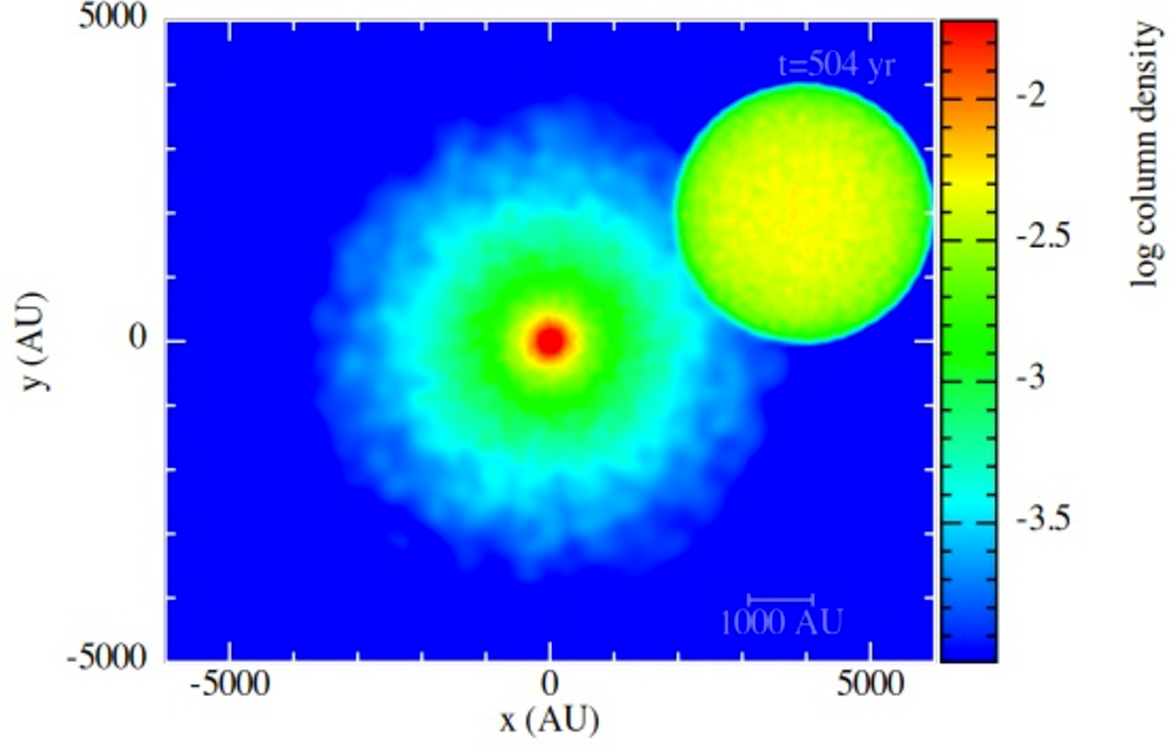}
    \includegraphics[width=3.3in]{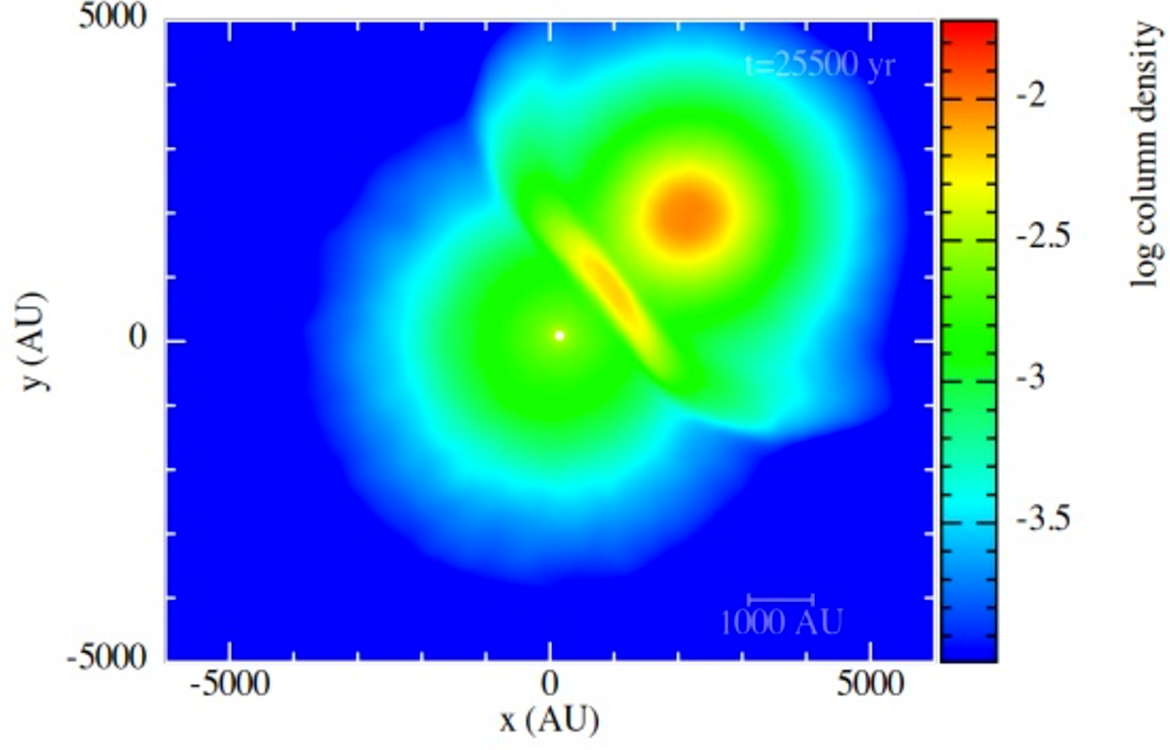}
    \includegraphics[width=3.3in]{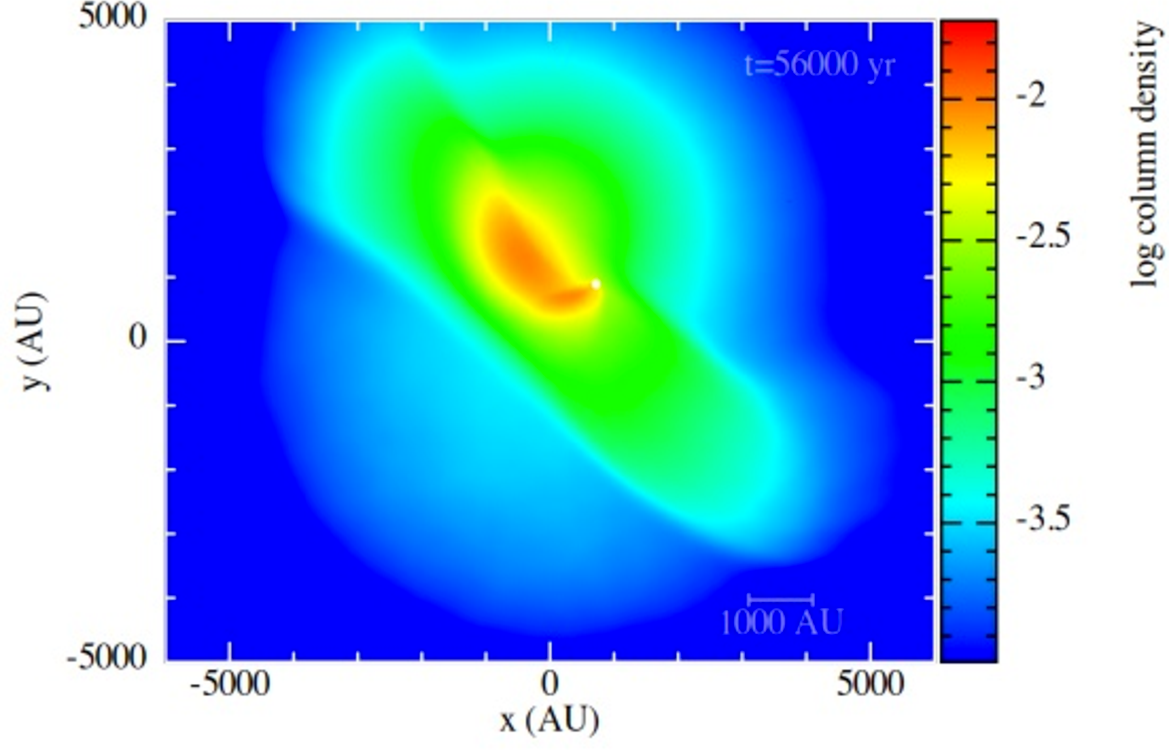}
           \caption{Cartoons for the twisted pseudo-disk (top panel) and core collision (middle and bottom panels) models, respectively. The core collision snapshots show the column density (in $\rm  g\ cm^{-2}$) at characteristic times during the evolution of the collision of the two cores. }
           \label{model}
  \end{figure*}

As the cloud core collapse progresses with time, the simulations produce the physical components of a pseudo-disk, a rotationally supported disk, and a jet/outflow, similar to the components produced in core collapse simulations when the initial rotation axis is aligned with the global magnetic field (Machida et al. 2009). However, the rotation and magnetic field are anisotropic forces and break the spherical symmetry in the collapsing cloud. Near the protostar, the magnetic field lines are strongly twisted by the rotational motion of the disk (Fig.~\ref{model}). The simulation results show that spiral arms appear on the size scale of $\geq$ 500 au and are connected to the central component. The arm-like structures extend to even $\sim$2000 au scale. At the large scale, we can see that a filamentary structure is formed from a pseudo-disk, in which the pseudo-disk is twisted by the rotational motion (Fig.~\ref{model}). The spiral arms are seen within the pseudo-disk, which is produced owing to the magnetic field and the core rotation. The material in the arms accretes toward the central region through the spiral arms. The size of the spiral arms in these models shows an increase with time. We have used the simulation data at the end of the simulations when the protostar has a mass of $\sim$20 M$_{Jup}$ at the epoch of $\sim$500 yr.

The full curved structure of the streamer cannot be reproduced by the twisted pseudo-disk model. The streamer in this model originates from the pseudo-disk and thus has an internal origin in the collapsing core, i.e., the streamer is not introduced from the external medium. Streamer-like structures thus appear universally in these simulations as long as the host cloud for brown dwarfs or low-mass stars is magnetized. However, the timescale for brown dwarf formation is as short as 10$^{4}$ years, which makes it difficult to observe a large-scale streamer in the brown dwarf formation process.

\subsection{Core collision model}

Structures like large spirals could form when a gas clump falls towards another cloud. Such collisions create elongated structures around young protostars/proto-brown dwarfs. We performed radiative hydrodynamic simulations of a gravitationally unstable core colliding with a gravitationally stable core (Fig.~\ref{model}). The unstable core collapses to form a bound proto-brown dwarf. The collision between the two cores forms a compressed layer of gas that wraps around the primary core, and creates a bright spiral-like structure (Fig.~\ref{model}). 

We used the SPH code SEREN (Hubber et al. 2011) to simulate the collision of two cores: (i) the {\it primary} core that is gravitationally unstable, collapses and forms a proto-brown dwarf, and (ii) the {\it secondary core} that is gravitationally stable. A sink (representing the bound proto-brown dwarf) is introduced when the density increases above $\rho_{\rm sink}=10^{-9} \textup{ g cm}^{-3}$. The sink interacts with the rest of the computational domain both gravitationally and radiatively; accretion luminosity dominates its emission. This radiation is taken into account assuming that gas accretion is continuous (Stamatellos et al. 2011; 2012). SPH particles which pass within $R_{\rm sink}=1$~AU and are gravitationally bound to a sink are accreted onto it.

The heating and cooling of gas is treated using the approximate radiative transfer method of Stamatellos et al. (2007) and Forgan et al. (2009), where the density and the gravitational potential of a gas particle are used to estimate a column density through which cooling/heating happens, and along with the local opacity, are used to estimate an optical depth for each particle. This is then used to determine the heating and cooling of the particle and includes effects from the rotational and vibrational degrees of freedom of $\textup{H}_{2}$, the dissociation of $\textup{H}_{2}$, ice melting, dust sublimation, bound-free, free-free and electron scattering interactions (Stamatellos et al. 2007).

We assume that the primary core has an initial density profile 
\begin{eqnarray}
\rho_{_{\rm P}}(r)&=&\frac{\rho_{_{\rm KERNEL}}}{\left(1\,+\,\left(r/R_{_{\rm KERNEL}}\right)^2\right)^{p/2}}\,.
\end{eqnarray}
where
$p$ defines the power-law for the decline at large distances,
$\rho_{_{\rm KERNEL}}$ is the central density, and $R_{_{\rm KERNEL}}$ is the radius of the central region within which the density is approximately uniform. The secondary core for simplicity is assumed to have uniform density.

We set $p=2.5$, $\rho_{_{\rm KERNEL}}=3\times 10^{-14}\,{\rm g}\,{\rm cm}^{-3}$, and $R_{_{\rm KERNEL}}=20\,{\rm AU}$. The core extends to $R_{\rm P}=5000\,{\rm AU}$, so its total mass is $M_{\rm P}=0.15\,{\rm M}_\odot$. The gas initial temperature is $T_{\rm P}=10\,{\rm K}$. We also include a mild uniform rotation of the core such that $\alpha_{_{\rm ROT}}\equiv{U_{_{\rm ROT}}}/{|U_{_{\rm GRAV}}|}=0.2$. The secondary core has size $R_{\rm S}=2,000$~AU, mass $M_{\rm S}=0.2\,{\rm M}_\odot$ and temperature $T_{\rm S}=10$~K. The secondary core is assumed to have a uniform density of $\rho_{\rm S}=3.6\times 10^{-17}\,{\rm g}\,{\rm cm}^{-3}$. The primary core is placed stationary at the origin of the coordinates, whereas the secondary core is placed at $(4000,2000,0)$~AU with a velocity on the $-x$ direction of $v_x=5~{\rm km\ s^{-1}}$. Therefore the impact parameter of the collision is $d=2000$~AU. The primary core is represented by $10^{6}$ SPH particles, so each SPH particle has mass $m_{_{\rm SPH}}\simeq 1.5\times 10^{-7}\,{\rm M}_\odot$. The minimum resolvable mass is therefore $M_{_{\rm MIN}}\simeq {\cal N}_{_{\rm NEIB}}m_{_{\rm SPH}}\simeq 7.5\times 10^{-6}\,{\rm M}_\odot$. The secondary core is represented by 1355030 SPH particles (so that the mass of each SPH particle representing the secondary core is the same as the mass of an SPH particle representing the primary core).

As seen in simulations of collisions on much larger-scale clouds (Whitworth et al. 1995; Kitsionas et al. 2007; Balfour et al. 2017), a compressed layer of gas forms at the interface between the two cores (see Fig.~\ref{model}). This compressed layer wraps around the primary core, which in the mean time has collapsed and formed a bound proto-brown dwarf (depicted by the white point) and creates a streamer (small structure) and spiral (large structure). The asymmetry is due to the collision not happening head-on, which is the most likely scenario within star forming regions. 

Faint streamer-like structures are also produced during the core collision simulations but are dispersed soon after formation. We can expect that the chaotic nature of core collision can give rise to such asymmetric structures. However, the colliding cloud cores model does not include accretion from an external medium, which may be associated with the large-scale streamer.

\subsection{Radiative transfer modelling}
\label{RTmodeling}

We have used the physical structure from the simulation data for the two models to conduct line radiative transfer modelling of the HCO$^{+}$ spectrum and compare the model moment 0 map with the observed map. The physical structure, i.e. the radial profiles of density, temperature, and velocity from the simulations is used as an input to the 3D non-LTE radiative transfer code MOLLIE (Keto et al. 2004). We generate synthetic HCO$^{+}$ spectra for different radial profiles of the molecular abundance. The synthetic line profiles are convolved by the ALMA beam size and compared with the observed blue-shifted spectrum for the spiral. A reasonable fit to the strength and width of the observed line profile is reached from varying the abundance profile and the line width. For each synthetic spectrum, a reduced-$\chi^{2}$ value is computed to determine the goodness of fit. Figure~4a shows the best model fit (lowest reduced-$\chi^{2}$ value) to the observed spectrum for the spiral for the twisted pseudo-disk and core collision models. From the best line model fit, we produced a moment 0 map to compare with the observed map. The HCO$^{+}$ molecular abundance relative to H$_{2}$ derived from both models is (0.8-2)$\times$10$^{-8}$ for the spiral. The model moment maps were generated by integrating the emission over the same velocity range. We estimate an uncertainty of $\sim$20\%-30\% on the molecular abundance.


\bsp	
\label{lastpage}
\end{document}